\documentclass[twocolumn,aps,prc,showpacs,superscriptaddress,floatfix]{revtex4}
\usepackage{epsfig}
\usepackage{color}
\usepackage{color}

\newcommand {\hic}	{{\sc hic}}
\newcommand {\rhic}	{{\sc rhic}}
\newcommand {\lhc}	{{\sc lhc}}
\newcommand {\cs}	{{\sc cs}}
\newcommand {\ws}	{{\sc ws}}
\newcommand {\dft}	{{\sc dft}}
\newcommand {\mcg}	{{\sc mcg}}
\newcommand {\ampt}	{{\sc ampt}}

\newcommand {\Au}	{$^{197}_{\;\,79}$Au}
\newcommand {\Cu}	{$^{62}_{29}$Cu}
\newcommand {\Pb}	{$^{207}_{\;\,82}$Pb}
\newcommand {\Ru}	{$^{96}_{44}$Ru}
\newcommand {\Zr}	{$^{96}_{40}$Zr}
\newcommand {\RuRu}	{RuRu}
\newcommand {\ZrZr}	{ZrZr}
\newcommand {\AuAu}	{AuAu}
\newcommand {\CuCu}	{CuCu}
\newcommand {\PbPb}	{PbPb}
\newcommand {\cme}	{{\mbox{\sc cme}}}
\newcommand {\bkg}	{{\mbox{\sc bkg}}}
\newcommand {\Qw}	{Q_{w}}
\newcommand {\Npart}	{N_{\rm part}}

\newcommand {\aEM}	{\alpha_{_{\rm EM}}}
\newcommand {\signn}	{\sigma_{_{\rm NN}}}
\newcommand {\dmin}	{d_{\rm min}}
\newcommand {\RP}	{{\sc rp}}
\newcommand {\PP}	{{\sc pp}}
\newcommand {\EP}	{{\sc ep}}
\newcommand {\tpc}	{{\sc tpc}}
\newcommand {\ftpc}	{{\sc ftpc}}
\newcommand {\zdc}	{{\sc zdc}}
\newcommand {\rcme}	{r}
\newcommand {\fcmeRP}	{f^{\rm RP}_{_{\cme}}}
\newcommand {\fcmeEP}	{f^{\rm EP}_{_{\cme}}}

\newcommand {\rt}	{r_{\perp i}}
\newcommand {\rpr}	{r^{\prime}}

\newcommand {\vv}	{v_{_{2}}}

\newcommand {\phia}	{\phi_{\alpha}}
\newcommand {\phib}	{\phi_{\beta}}

\newcommand {\psiPPEP}	{\psi_{_{\rm PP(EP)}}}
\newcommand {\psione}	{\psi_{_{1}}}
\newcommand {\psitwo}	{\psi_{_{2}}}
\newcommand {\psiRP}	{\psi_{_{\rm RP}}}
\newcommand {\psiPP}	{\psi_{_{\rm PP}}}
\newcommand {\psiEP}	{\psi_{_{\rm EP}}}
\newcommand {\psiB}	{\psi_{_{\rm B}}}
\newcommand {\dpsi}	{\mean{\cos2(\psiPP-\psiRP)}}
\newcommand {\dpsiEP}	{\mean{\cos2(\psiEP-\psiRP)}/\res}
\newcommand {\etwo}	{\epsilon_{_{2}}}

\newcommand {\eRP}	{\etwo\{\psiRP\}}
\newcommand {\ePP}	{\etwo\{\psiPP\}}
\newcommand {\vpsi}	{\vv\{\psi\}}
\newcommand {\vRP}	{\vv\{\psiRP\}}

\newcommand {\vEP}	{\vv\{\psiEP\}}
\newcommand {\res}	{\mathcal{R}_{\rm EP}}
\newcommand {\vTT}	{\vv\{2\}}
\newcommand {\vTF}	{\vv\{4\}}
\newcommand {\vtpc}	{\vv^{\mbox\tpc}}
\newcommand {\vftpc}	{\vv^{\mbox\ftpc}}
\newcommand {\vzdc}	{\vv^{\mbox\zdc}}

\newcommand {\aee}	{a_{\etwo}^{\rm PP}}
\newcommand {\av}	{a_{\vv}^{\rm EP}}
\newcommand {\aB}	{a_{_{\Bsq}}}
\newcommand {\aPP}	{a^{\rm PP}}
\newcommand {\aEP}	{a^{\rm EP}}
\newcommand {\aBPP}	{\aB^{\rm PP}}
\newcommand {\aBEP}	{\aB^{\rm EP}}
\newcommand {\RPP}	{R^{\rm PP}}
\newcommand {\REP}	{R^{\rm EP}}
\newcommand {\Rexp}	{R^{\rm exp}}
\newcommand {\rPP}	{R_{\rm PP}}
\newcommand {\rEP}	{R_{\rm EP}}
\newcommand {\rPPEP}	{R_{\rm PP(EP)}}
\newcommand {\RPPEP}	{R^{\rm PP(EP)}}
\newcommand {\Bcos}	{\mean{(eB(\mathbf{0},0)/m_\pi^2)^2\cos2(\psiB-\psi)}}

\newcommand {\BcosEP}	{\mean{(eB(\mathbf{0},0)/m_\pi^2)^2\cos2(\psiB-\psiEP)}/\res}
\newcommand {\Bbf}	{$\mathbf{B}$}
\newcommand {\Bsq}	{B_{\rm sq}}
\newcommand {\Bpsi}	{\Bsq\{\psi\}}
\newcommand {\BRP}	{\Bsq\{\psiRP\}}
\newcommand {\BPP}	{\Bsq\{\psiPP\}}
\newcommand {\BEP}	{\Bsq\{\psiEP\}}
\newcommand {\gOS}	{\gamma_{_{\rm OS}}}
\newcommand {\gSS}	{\gamma_{_{\rm SS}}}
\newcommand {\dg}	{\Delta\gamma}
\newcommand {\dgpsi}	{\dg\{\psi\}}
\newcommand {\dgRP}	{\dg\{\psiRP\}}

\newcommand {\dgEP}	{\dg\{\psiEP\}}
\newcommand {\dgone}	{\dg\{\psione\}}
\newcommand {\dgtwo}	{\dg\{\psitwo\}}

\newcommand {\mean}[1]	{\langle #1\rangle}

\begin{document}
\title{Varying the chiral magnetic effect relative to flow in a single nucleus-nucleus collision}
\author{Hao-jie Xu}
\address{School of Science, Huzhou University, Huzhou, Zhejiang 313000, China}
\author{Jie Zhao}
\address{Department of Physics and Astronomy, Purdue University, West Lafayette, Indiana 47907, USA}
\author{Xiaobao Wang}
\address{School of Science, Huzhou University, Huzhou, Zhejiang 313000, China}
\author{Hanlin Li}
\address{College of Science, Wuhan University of Science and Technology, Wuhan, Hubei 430065, China}
\author{Zi-Wei Lin}
\address{Department of Physics, East Carolina University, Greenville, North Carolina 27858, USA}
\address{Key Laboratory of Quarks and Lepton Physics (MOE) and Institute of Particle Physics, Central China Normal University, Wuhan, Hubei 430079, China}
\author{Caiwan Shen}
\address{School of Science, Huzhou University, Huzhou, Zhejiang 313000, China}
\author{Fuqiang Wang}
\address{School of Science, Huzhou University, Huzhou, Zhejiang 313000, China}
\address{Department of Physics and Astronomy, Purdue University, West Lafayette, Indiana 47907, USA}
\date{\today}

\begin{abstract}
We propose a novel method to search for the chiral magnetic effect (\cme) in heavy ion collisions. We argue that the relative strength of the magnetic field (mainly from spectator protons and responsible for the \cme) with respect to the reaction plane and the participant plane is opposite to that of the elliptic flow background arising from the fluctuating participant geometry. This opposite behavior in a single collision system, hence with small systematic uncertainties, 
can be exploited to extract the possible \cme\ signal from the flow background. The method is applied to the existing data at \rhic, the outcome of which is discussed.
\end{abstract}



\maketitle
\section{Introduction}
One of the fundamental properties of quantum chromodynamics (QCD) is the creation of topological gluon fields from vacuum fluctuations in local domains~\cite{Lee:1974ma,Morley:1983wr,Kharzeev:1998kz}. 
The interactions with those gluon fields can change the handedness (chirality) of quarks, under restoration of the approximate chiral ($\chi$) symmetry, which would lead to local parity and charge-conjugate parity violations~\cite{Kharzeev:1998kz,Kharzeev:1999cz,Kharzeev:2007jp}. These violations in the early universe could be responsible for the matter-antimatter asymmetry~\cite{RevModPhys.76.1} and are of fundamental importance.
The resulting chirality imbalance, often referred to as the topological charge ($\Qw$), can lead to an electric current, or charge separation (\cs), along a strong magnetic field (\Bbf), a phenomenon called the chiral magnetic effect (\cme)~\cite{Fukushima:2008xe}. 
Such phenomena 
are not specific to QCD but 
a subject of interest to a wide range of physics communities~\cite{Kharzeev:2015znc}, 
e.g.~condensed matter and materials physics~\cite{Li:2014bha,Lv:2015pya,Huang:2015eia}.

\begin{figure}
	\begin{center}
    \includegraphics[width=0.4\textwidth]{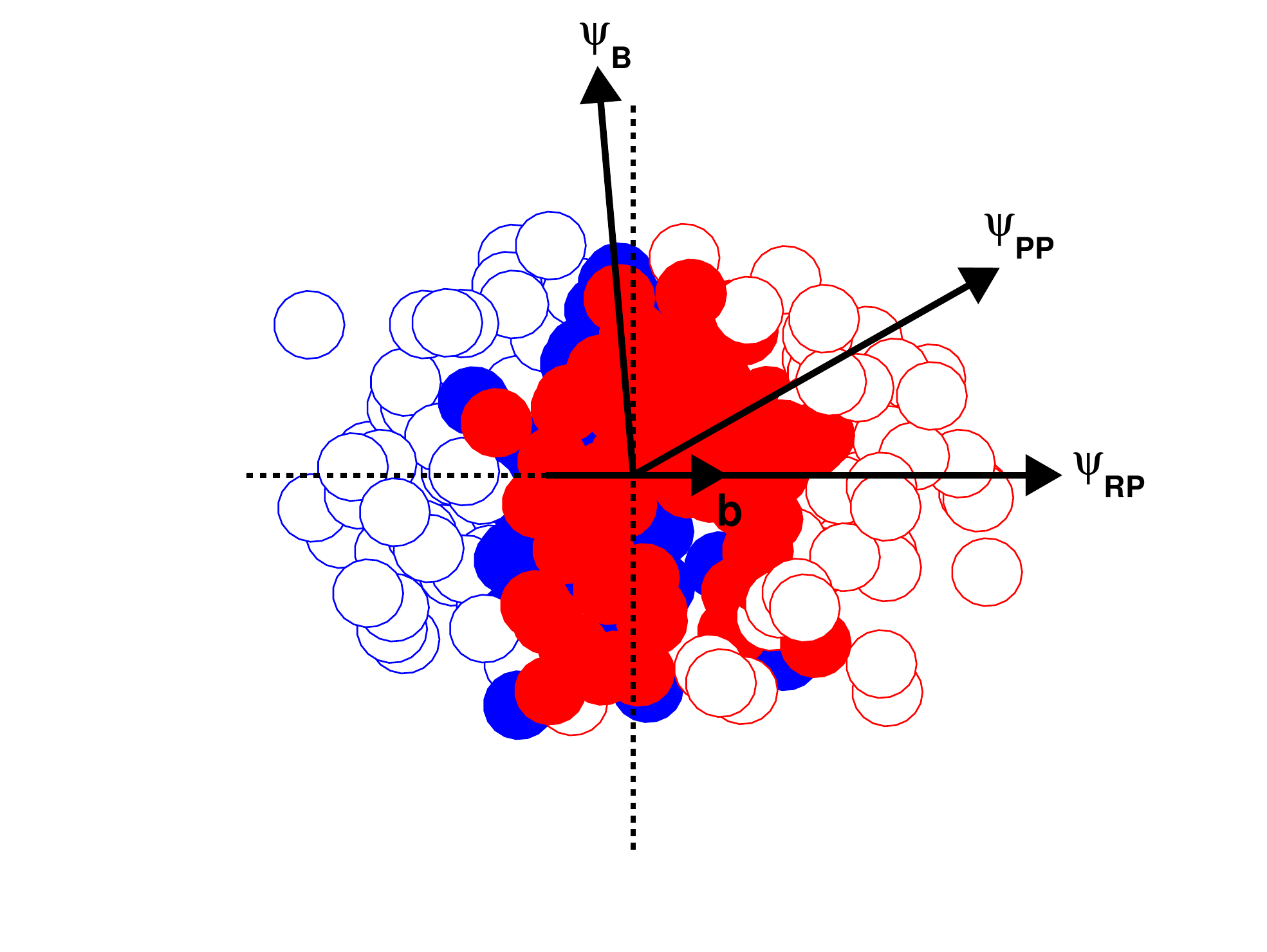}
  \caption{\label{sketch}(Color online) Sketch of a heavy ion collision projected onto the transverse plane (perpendicular to the beam direction). $\psiRP$ is the reaction plane (impact parameter, $\mathbf{b}$) direction, $\psiPP$ the participant plane direction (of interacting nucleons, denoted by the solid circles), and $\psiB$ the magnetic field direction (mainly from spectator protons, denoted by the open circles together with spectator neutrons).}
  
  \end{center}
\end{figure}

Both conditions for the \cme--the $\chi$-symmetry and a strong \Bbf~\cite{Kharzeev:2015znc}--may be met in relativistic heavy ion collisions (\hic), where a high energy-density state of deconfined quarks and gluons (the quark-gluon plasma) is formed, resembling the conditions of the early universe~\cite{Arsene:2004fa,Adcox:2004mh,Back:2004je,Adams:2005dq,Muller:2012zq}. 
At high energies, a nucleus-nucleus collision can be considered as being composed of participant nucleons in an overlap interaction zone and the rest spectator nucleons passing by continuing into the beam line (see the sketch of a nucleus-nucleus collision in Fig.~\ref{sketch}). The spectator protons produce the majority of \Bbf, whose direction is, on average, perpendicular to the reaction plane (\RP) spanned by the impact parameter ($\mathbf{b}$) and beam directions. 
The sign of the $\Qw$ is, however, 
random due to its fluctuating nature; consequently the \cs\ dipole direction 
is random~\cite{Kharzeev:2007jp}. 
As a result the \cs\ can only be measured by charge correlations. 
A commonly used observable is the three-point azimuthal correlator~\cite{Voloshin:2004vk},
$\gamma\equiv\mean{\cos(\phi_\alpha+\phi_\beta-2\psiRP)}$,
where $\phi_\alpha$ and $\phi_\beta$ are the azimuths of two charged particles, and $\psiRP$ is that of the \RP.
Because of charge-independent backgrounds, such as correlations from global momentum conservation, 
the correlator difference between opposite-sign ({\sc os}) and same-sign ({\sc ss}) pairs, $\dg\equiv\gOS-\gSS$, is used. 
However, $\dg$ is ambiguous between a back-to-back pair perpendicular to the \RP\ (potential \cme\ signal) and an aligned pair in the \RP\ (e.g.~from resonance decay). There are generally more particles (including resonances) produced along the \RP\ than perpendicular to it, the magnitude of which is characterized by the elliptic anisotropy parameter ($\vv$)~\cite{Reisdorf:1997fx}. It is commonly interpreted as coming from a stronger hydrodynamic push in the short-axis (i.e.~\RP) direction of the elliptically-shaped overlap zone between the two colliding nuclei~\cite{Ollitrault:1992bk}. As a result, $\dg$ is contaminated by a background~\cite{Voloshin:2004vk,Wang:2009kd,Bzdak:2009fc,Schlichting:2010qia,Wang:2016iov,Zhao:2018ixy}, which arises from the coupling between particle correlations and $\vv$, and is hence proportional to $\vv$.

  The search for the \cme\ is one of the most active research in \hic\ at the Relativistic Heavy Ion Collider (\rhic) and the Large Hadron Collider (\lhc)~\cite{Abelev:2009ac,Abelev:2009ad,Abelev:2012pa,Adamczyk:2013hsi,Adamczyk:2014mzf,Adamczyk:2013kcb,Khachatryan:2016got,Sirunyan:2017quh,Acharya:2017fau}. 
A finite $\dg$ signal is observed~\cite{Abelev:2009ac,Abelev:2009ad,Adamczyk:2014mzf,Adamczyk:2013hsi,Abelev:2012pa}, but how much background contamination is not yet settled.
There have been many attempts to gauge, reduce or eliminate the flow backgrounds, by event-by-event $\vv$ dependence~\cite{Adamczyk:2013kcb}, 
event-shape engineering~\cite{Sirunyan:2017quh,Acharya:2017fau}, 
comparing to small-system collisions~\cite{Zhao:2017wck,Khachatryan:2016got,Sirunyan:2017quh}, invariant mass study~\cite{Zhao:2017nfq}, and by new observables~\cite{Ajitanand:2010rc,Magdy:2017yje}. 
The \lhc\ data seem to suggest that the \cme\ signal is small and consistent with zero~\cite{Sirunyan:2017quh,Acharya:2017fau}, while the situation at \rhic\ is less clear~\cite{Kharzeev:2015znc}. 

To better gauge background contributions, isobaric \Ru+\Ru\ (\RuRu) and \Zr+\Zr\ (\ZrZr) collisions have been proposed~\cite{Voloshin:2010ut} and planned at \rhic\ in 2018. 
Their QCD backgrounds are expected to be almost the same because of the same mass number, whereas the atomic numbers, hence \Bbf, differ by 10\%. These expectations are qualitatively confirmed by studies~\cite{Deng:2016knn} with Woods-Saxon (\ws) nuclear densities; the \cme\ signal over background could be improved by a factor of seven in comparative measurements of \RuRu\ and \ZrZr\ collisions than each of them individually.
A recent study by us~\cite{Xu:2017zcn} has shown, however, that there could exist large uncertainties on the differences in both the overlap geometry eccentricity ($\etwo$) and \Bbf\ 
due to nuclear density deviations 
from \ws. As a result, the isobaric collisions may not provide a clear-cut answer to the existence or the lack of the \cme.

In what follows, we argue that one has, in a single collision system, all and {\em even better} advantages of significant \Bbf\ and minimal $\etwo$ differences of the comparative isobaric collisions, {\em and} with the benefit of minimal theoretical and experimental uncertainties. The idea is straightforward, as illustrated in Fig.~\ref{sketch}. \Bbf\ is produced by spectator protons hence its projection, on average, is the strongest perpendicular to the \RP~\cite{Kharzeev:2007jp}; $\vv$ stems from the collision geometry and is the strongest with respect to the second harmonic participant plane (\PP)~\cite{Alver:2006wh}. The \RP\ and the \PP\ are correlated but, due to fluctuations~\cite{Alver:2006wh}, not identical. Measurements with respect to the \RP\ and the \PP, therefore, contain different amounts of \cme\ signal and $\vv$ background, and thus can help disentangle the two contributions.

\section{General idea}
Due to fluctuations, the \PP\ azimuthal angle ($\psiPP$) is not necessarily aligned with the \RP's~\cite{Alver:2006wh}. 
The $\vv$ is directly related to the eccentricity of the transverse overlap geometry, $\ePP\equiv\mean{\ePP_{\rm evt}}$.
The average is taken over the event-by-event eccentricity magnitudes, which can be obtained by~\cite{Miller:2007ri,Alver:2006wh,Alver:2008zza,Zhu:2016puf,Xu:2017zcn} 
\begin{equation}
\ePP_{\rm evt}e^{i2\psiPP}=\sum_{i=1}^{\Npart}\left(\rt^{2}e^{i2\phi_{\rt}}\right)\left/\sum_{i=1}^{\Npart}\rt^{2}\right.\,,
\label{eq:ecc}
\end{equation}
where $(\rt,\phi_{\rt})$ is the polar coordinate of the $i$-th participant nucleon. 
The overlap geometry relative to $\mathbf{b}$, averaged over many events, is an ellipse with its short axis being along the \RP; its eccentricity is
\begin{equation}
\eRP=\mean{\ePP_{\rm evt}\cos2(\psiPP-\psiRP)}\,. 
\label{eq:eRP}
\end{equation}
Let 
\begin{equation}
\aPP\equiv\dpsi\;
\label{eq:a}
\end{equation}
measure the correlation between $\psiPP$ and $\psiRP$. We have
\begin{equation}
\aee\equiv\eRP/\ePP\approx\aPP\;.
\label{eq:aee}
\end{equation}
The factorization is approximate, valid only when, at a given collision centrality, the $\ePP_{\rm evt}$ magnitude does not vary with the $\psiPP$ fluctuation around $\psiRP$. 

\Bbf\ is mainly produced by spectator protons. Their positions fluctuate; the \Bbf\ azimuthal direction, $\psiB$, is not always perpendicular to the \RP~\cite{Bzdak:2011yy,Deng:2012pc,Bloczynski:2012en} (see illustration in Fig.~\ref{sketch}). The \cme-induced \cs\ is along the \Bbf\ direction~\cite{Kharzeev:2007jp}; 
when measured perpendicular to a direction $\psi$, its relevant strength is proportional to $\Bpsi\equiv\Bcos$~\cite{Deng:2016knn}. 
Although the field point $\mathbf{r}=0$ is used here, other field points are also calculated and our conclusion does not change.
The overall magnetic field strength is calculated at $t=0$. In general, the magnetic field changes when the system evolves, and there are large theoretical uncertainties. However, as we will show that only the relative difference is used in our method, the absolute magnitudes of the magnetic field do not affect our conclusions.
Because the position fluctuations of participant nucleons and spectator protons are independent except the overall constrain of the nucleus, $\psiPP$ and $\psiB$ fluctuate independently about $\psiRP$. 
This yields 
\begin{equation}
\aBPP\equiv\BPP/\BRP\approx\aPP\;.
\label{eq:aB}
\end{equation}
Because \Bbf\ contains contributions also from participant protons, 
the factorization is only approximate.
We calculate $B(\mathbf{r},t)$ by~\cite{Bzdak:2011yy,Deng:2012pc,Bloczynski:2012en}
\begin{equation}
	e\mathbf{B}(\mathbf{r},t)=\sum_{i}\aEM Z_{i}(\mathbf{\rpr_{i}})\frac{1-v_{i}^{2}}{[r_{i}^{\prime\,2}-(\mathbf{\rpr_{i}}\times\mathbf{v_{i}})^{2}]^{3/2}}\mathbf{v_{i}}\times\mathbf{\rpr_{i}}\,,
\end{equation}
where $\mathbf{v_{i}}$ is the velocity of the $i$-th proton, $\mathbf{\rpr_{i}} = \mathbf{r} -\mathbf{r_{i}}(t)$ is the relative distance between the field point $\mathbf{r}$ and the proton position $\mathbf{r_{i}}(t)$ at time $t$, $\aEM=1/137$, and $Z_{i}(\mathbf{\rpr_{i}})$ is the charge number factor. After employing a finite proton radius  $r_{p}= 0.88$ fm, $Z_{i}(\mathbf{\rpr_{i}})=1$ if $\mathbf{r}$ locates outside the proton and $Z_{i}(\mathbf{\rpr_{i}})<1$ depends on $\mathbf{\rpr_{i}}$ if $\mathbf{r}$ locates inside the proton~\cite{Bloczynski:2012en}. Varying the $r_{p}$ value has little effect on our results.

\begin{figure*}
  \begin{center}
    \includegraphics[width=0.9\textwidth]{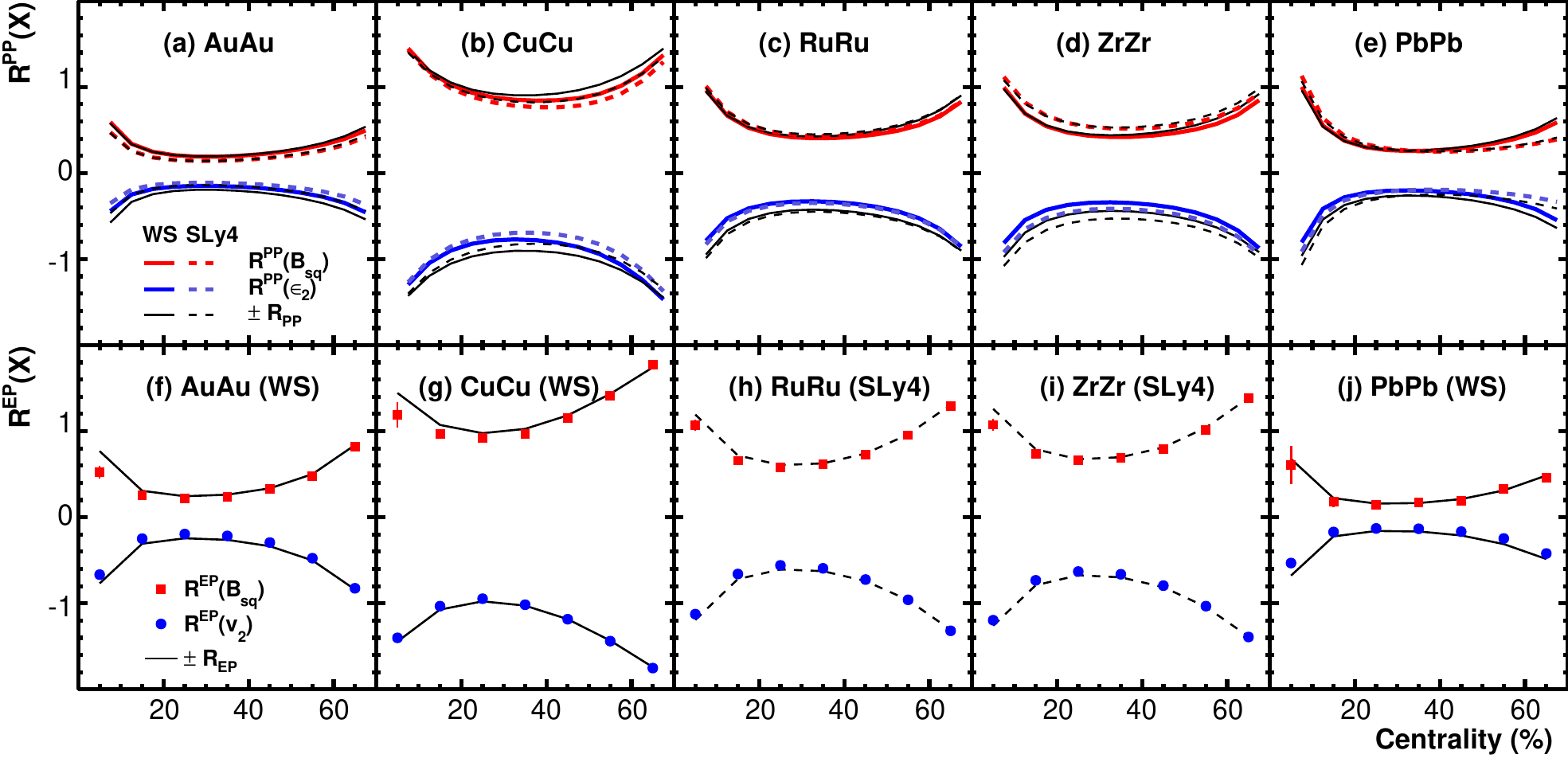}
  \caption{\label{fig}(Color online) Relative differences $\RPP(\etwo)$, $\RPP(\Bsq)$, $\RPP$ from \mcg\ (upper panel) and $\REP(\vv)$, $\REP(\Bsq)$, $\REP$ from \ampt\ (lower panel) for (a,f) \AuAu, (b,g) \CuCu, (c,h) \RuRu, and (d,i) \ZrZr\ at \rhic, and (e,j) \PbPb\ at the \lhc. Both the \ws\ and \dft-calculated densities are shown for the \mcg\ results, while the used density profiles are noted for the \ampt\ results. Errors, mostly smaller than the symbol size, are statistical.}
  \end{center}
\end{figure*}

It is convenient to define a relative difference~\cite{Deng:2016knn},
\begin{equation}
\RPPEP(X)\equiv2\cdot\frac{X\{\psiRP\}-X\{\psiPPEP\}}{X\{\psiRP\}+X\{\psiPPEP\}}\;,
\end{equation}
where $X\{\psiRP\}$ and $X\{\psiPPEP\}$ are the measurements of quantity $X$ with respect to $\psiRP$ and $\psiPP$ (or $\psiEP$ described below), respectively. 
Those in $\etwo$ and $\Bsq$ are
\begin{eqnarray}
\RPP(\etwo)&\equiv&-2(1-\aee)/(1+\aee)\approx-\rPP\;,\nonumber\\
\RPP(\Bsq)&\equiv&2(1-\aBPP)/(1+\aBPP)\approx\rPP\;,
\label{eq:R}
\end{eqnarray}
where
\begin{equation}
\rPP\equiv2(1-\aPP)/(1+\aPP)\;.
\end{equation}
The upper panels of Fig.~\ref{fig} show $\RPP(\etwo)$ and $\RPP(\Bsq)$ calculated by a {\em Monte Carlo} Glauber model (\mcg)~\cite{Miller:2007ri,Alver:2008zza,Xu:2014ada,Zhu:2016puf} for \Au+\Au\ (\AuAu), \Cu+\Cu\ (\CuCu), \RuRu, \ZrZr\ collisions at \rhic\ and \Pb+\Pb\ (\PbPb) collisions at the \lhc. 
The centrality percentage is determined from the impact parameter, $b=|\mathbf{b}|$, in \mcg.
For a given $b$ drawn from the probability distribution $P(b)\propto b$, a participant nucleon is determined by its relative transverse distance $d$ from the surrounding nucleons in the other nucleus, according to the nucleon-nucleon differential cross section $d^2\signn/d^2d=A\exp(-\pi Ad^{2}/\signn)$~\cite{Rybczynski:2011wv}. We use $A=0.92$; $\signn = 42$~mb  for \AuAu, \CuCu, \RuRu\ and \ZrZr\ collisions and $\signn=62.4$~mb for \PbPb\ collisions. 
The minimum inter-nucleon distance in each nucleus is set to be $\dmin=0.4$~fm. 
Varying $\signn$ and $\dmin$ does not change our results significantly.
At the same energy (\rhic), the smaller the system, the larger the fluctuations and hence the larger the $\RPP(\etwo)$ and $\RPP(\Bsq)$ magnitudes. The larger \lhc\ value for \PbPb\ than the \rhic\ value for \AuAu\ is due to the larger nucleon-nucleon cross section ($\signn$) used at \lhc\ than \rhic.
Spherical nuclei with the \ws\ 
as well as the energy density functional theory (\dft)~\cite{Bender:2003jk,Erler:2012xxx} calculated distributions~\cite{Xu:2017zcn} are used. The uncertainties in the \dft\ calculations, assessed by using different mean fields (SLy4 and SLy5~\cite{Chabanat:1997un}, and SkM*~\cite{Bartel:1982ed}), with (Hartree-Fock-Bogoliubov) and without (Hartree-Fock) pairing correlations~\cite{Bender:2003jk,ring2000nuclear,Wang:2016rqh}, and varying the nuclear deformations~\cite{Xu:2017zcn}, 
are all small on our results. 
The results are compared to the corresponding $\pm\rPP$: 
the approximations in Eq.~(\ref{eq:aee}) and (\ref{eq:aB}) are good. 
	  The \PP\ is not experimentally measured, nor is $\etwo$. As a proxy for \PP, the event plane (\EP) is often reconstructed by $\vEP_{\rm evt}e^{i2\psiEP}=\mean{e^{i2\phi}}$, similar to Eq.~(\ref{eq:ecc}), but using final-state charged particle azimuthal angle $\phi$ in momentum space~\cite{Poskanzer:1998yz}. 

The $\vv$ is measured by the \EP\ method with a correction for the \EP\ resolution ($\res$), 
$\vEP=\mean{\cos2(\phi-\psiEP)}/\res$, 
where $\phi$ is the particle momentum azimuthal angle, or almost equivalently, by two-particle correlations,
$\vEP\approx \vTT\equiv\mean{\cos2(\phia-\phib)}^{1/2}$~\cite{Poskanzer:1998yz}.
The \EP\ resolution $\res$ is calculated by the subevent method with an iterative procedure, dividing the particles randomly into two subevents~\cite{Poskanzer:1998yz}.
A $\vv$ can also be obtained with respect to the \RP, 
$\vRP\equiv\mean{\cos2(\phi-\psiRP)}$.
Although a theoretical concept, the \RP\ may be assessed by Zero-Degree Calorimeters (\zdc) measuring sidewards-kicked spectator neutrons (directed flow $v_1$)~\cite{Reisdorf:1997fx,Abelev:2013cva,Adamczyk:2016eux}.
Similar to Eqs.~(\ref{eq:a},\ref{eq:aee}), let
\begin{equation}
\aEP=\dpsiEP\;, 
\end{equation}
and we have
\begin{equation}
\av\equiv\vRP/\vEP\approx\aEP\;.
\end{equation}
As $\BPP$, one can obtain 
$\BEP=\BcosEP$,
and a similar relationship to Eq.~(\ref{eq:aB}),
\begin{equation}
\aBEP\equiv\BEP/\BRP\approx\aEP\;.
\end{equation}

The relative differences in $\vv$ and $\Bsq$ are
\begin{eqnarray}
\REP(\vv)&\equiv&-2(1-\av)/(\av+1)\approx-\rEP\;,\nonumber\\
\REP(\Bsq)&\equiv&2(1-\aBEP)/(1+\aBEP)\approx\rEP\;,
\end{eqnarray}
where
\begin{equation}
\rEP\equiv2(1-\aEP)/(1+\aEP)\;.
\end{equation}

The lower panels of Fig.~\ref{fig} show A Multi-Phase Transport (\ampt, ``string melting'') simulation results of $\REP(\vv)$ and $\REP(\Bsq)$, compared to $\pm\rEP$. 
Again, good agreements are found.
The \ampt\ centrality is determined from the midrapidity ($|\eta|<1$) final-state charged particle multiplicity~\cite{Xu:2017zcn}, similar to experiments~\cite{Abelev:2008ab}. Details of \ampt\ can be found in Refs.~\cite{Lin:2001zk,Lin:2004en,Lin:2014tya,Ma:2011uma}. The \ampt\ version (v2.26t5) and parameter values used in the present work are the same as those used earlier for \rhic\ collisions in~\cite{Lin:2014tya,Ma:2016fve,He:2015hfa,Li:2016flp,Li:2016ubw,Zhao:2017nfq}; the same parameter setting is used for \lhc\ energy. 
The \mcg\ and \ampt\ results cannot be readily compared quantitatively because the former involves \PP\ while the latter uses \EP\ as it would be in experiments. Although \ampt\ employs \mcg\ as its initial geometry, 
the subsequent parton-parton scatterings in \ampt\ are important for the final-state \EP\ determination.
In addition, other distinctions exist, such as the nuclear shadowing effect and the Gaussian implementation of $\signn$, which yield different predictions for the eccentricity (hence flow harmonics) and its fluctuations~\cite{Zhu:2016puf,Xu:2016hmp}. Nevertheless, the general features are similar between the \mcg\ and \ampt\ results. Both show the opposite behavior of $\RPPEP(\etwo(\vv))$ and $\RPPEP(\Bsq)$, which approximately equal to $\pm\rPPEP$.

The commonly used $\dg$ variable contains, in addition to the \cme\ it is designed for, $\vv$-induced background, 
\begin{equation}
\dgpsi=\cme(\Bpsi)+\bkg(\vpsi)\;.
\end{equation}
$\dgpsi$ can be measured with respect to $\psi=\psiRP$ (using the 1st order event plane $\psione$ by the \zdc) and $\psi=\psiEP$ (2nd order event plane $\psitwo$ via final-state particles). 
If $\bkg(\vv)$ is proportional to $\vv$~\cite{Voloshin:2004vk,Wang:2009kd,Bzdak:2009fc,Schlichting:2010qia,Wang:2016iov,Zhao:2018ixy} and $\cme(\Bsq)$ to $\Bsq$~\cite{Bloczynski:2012en}, then 
\begin{equation}
\REP(\dg)=2\frac{r(1-\aBEP)-(1-\av)}{r(1+\aBEP)+(1+\av)}\approx\frac{1-r}{1+r}\REP(\vv)\;.
\end{equation}
Here $\rcme\equiv\cme(\BRP)/\bkg(\vEP)$ can be considered as 
the relative \cme\ signal to background contribution,
\begin{equation}
\rcme
=\frac{1+\av}{1+\aBEP}\frac{\REP(\dg)-\REP(\vv)}{\REP(\Bsq)-\REP(\dg)}
\approx\frac{\REP(\vv)-\REP(\dg)}{\REP(\vv)+\REP(\dg)}\;.
\label{eq:r}
\end{equation}
If the experimental measurement $\REP(\dg)$ equals to $\REP(\vv)$ (i.e.~$\dg$ scales like $\vv$), then \cme\ contribution is zero; if $\REP(\dg)\approx-\REP(\vv)$ (i.e.~$\dg$ scales like $\Bsq$), then background is close to zero and all would be \cme; and if $R(\dg)=0$, then background and \cme\ contributions are of similar magnitudes.
The \cme\ signal fractions with respect to \RP\ and \EP\ are, respectively, 
\begin{eqnarray}
\fcmeRP&=&\cme(\BRP)/\dgRP=\rcme/(\rcme+\av)\;,\nonumber\\
\fcmeEP&=&\cme(\BEP)/\dgEP=\rcme/(\rcme+1/\aBEP)\;.
\end{eqnarray}

\section{Apply to data}
The quantities 
$\aPP$ and $\aEP$, and consequently $\RPP$ and $\REP$,
are mainly determined by fluctuations. 
Being defined in a single nucleus-nucleus collision, they are insensitive to many details, such as the structure functions of the colliding nuclei. 
This is in contrast to comparisons between two isobaric collision systems where large theoretical uncertainties are present~\cite{Xu:2017zcn}. There have been tremendous progresses over the past decade in our understanding of the nuclear collision geometry and fluctuations~\cite{Heinz:2013th}. The \mcg\ and \ampt\ calculations of these quantities are therefore on a rather firm ground. 

Experimentally, $\REP(\vv)$ can be assessed by $\vv$ measurements. $\REP(\Bsq)$ cannot but may be approximated by $-\REP(\vv)$, as demonstrated by the \mcg\ and \ampt\ calculations.
Table~\ref{tab:data} shows the measured $\vv$ in 200A~GeV \AuAu\ collisions by STAR via the \zdc\ $\psione$ at beam rapidities ($\vzdc$)~\cite{Wang:2005ab} and the forward time projection chamber (\ftpc) $\psitwo$ (i.e.~$\psiEP$) at forward/backward rapidities ($\vftpc$)~\cite{Voloshin:2007af}, together with those via the midrapidity \tpc\ \EP\ ($\vtpc$) and the two- and four-particle cumulants ($\vTT$, $\vTF$)~\cite{Adams:2004bi}. 
The relative difference ($\Rexp(\vv)$) between $\vzdc$ and $\vftpc$ is smaller in magnitude than $\RPP(\etwo)$ from \mcg\ and $\REP(\vv)$ from \ampt; moreover, $\vftpc$ may already be on the too-large side as it is larger than $\vtpc$ for some of the centralities whereas the opposite is expected because of a smaller nonflow contribution to $\vftpc$~\cite{Heinz:2013th,Abdelwahab:2014sge}.
These may suggest that $\vzdc$ may not measure the $\vv$ purely relative to the \RP, but a mixture of \RP\ and \PP. This is possible because, for instance, the \zdc\ could intercept not only spectator neutrons but also those having suffered only small-angle elastic scatterings.

\begin{table*}
\begin{center}
\caption{\label{tab:data}STAR midrapidity ($|\eta|<1$) charged particle $\vv$ (in percent) in 200A~GeV \AuAu\ collisions as functions of centrality (number of participants $\Npart$ from Ref.~\cite{Abelev:2008ab}): $\vzdc$~\cite{Wang:2005ab}, $\vftpc$~\cite{Voloshin:2007af}, and $\Rexp(\vv)\equiv2(\vzdc-\vftpc)/(\vzdc+\vftpc)$ compared to $\RPP(\etwo)$ from \mcg\ and $\REP(\vv)$ from \ampt. Also listed are $\vtpc$, $\vTT$, and $\vTF$~\cite{Adams:2004bi}, followed by the three-point correlator $\dgtwo$ and $\dgone$ measurements~\cite{Abelev:2009ac,Abelev:2009ad,Adamczyk:2013hsi}.}

\begin{tabular}{cc|ccc|cc|ccc|cc}\hline
cent.  &$\Npart$& $\vzdc$&$\vftpc$&$\Rexp(\vv)$ &$\RPP(\etwo)$ & $\REP(\vv)$ & $\vtpc$ & $\vTT$ & $\vTF$ & $\dgtwo\times10^4$ & $\dgone\times10^4$ \\\hline
50-60\%& 49.3  & 5.90 & 7.15 & $-0.19$ & $-0.240$ & $-0.478$ & 7.30 & 7.59 & 6.18 & $5.133\pm0.061$ & $5.07\pm1.60$\\
40-50\%& 78.3  & 6.40 & 7.34 & $-0.14$ & $-0.176$ & $-0.285$ & 7.25 & 7.64 & 6.43 & $3.393\pm0.026$ & $3.99\pm0.77$\\
30-40\%& 117.1 & 6.25 & 7.00 & $-0.11$ & $-0.145$ & $-0.219$ & 6.92 & 7.29 & 6.33 & $2.248\pm0.012$ & $2.18\pm0.44$ \\
20-30\%& 167.6 & 5.70 & 6.17 & $-0.08$ & $-0.147$ & $-0.197$ & 6.09 & 6.42 & 5.66 & $1.424\pm0.007$ & $1.56\pm0.31$ \\
20-60\%&       & 5.94 & 6.59 & $-0.10$ & $-0.155$ & $-0.227$ & 6.52 & 6.86 & 5.96 & $2.067\pm0.007$ & $2.19\pm0.24$ \\
\hline
\end{tabular}
\end{center}
\end{table*}

\begin{figure}
  \begin{center}
    \includegraphics[width=0.4\textwidth]{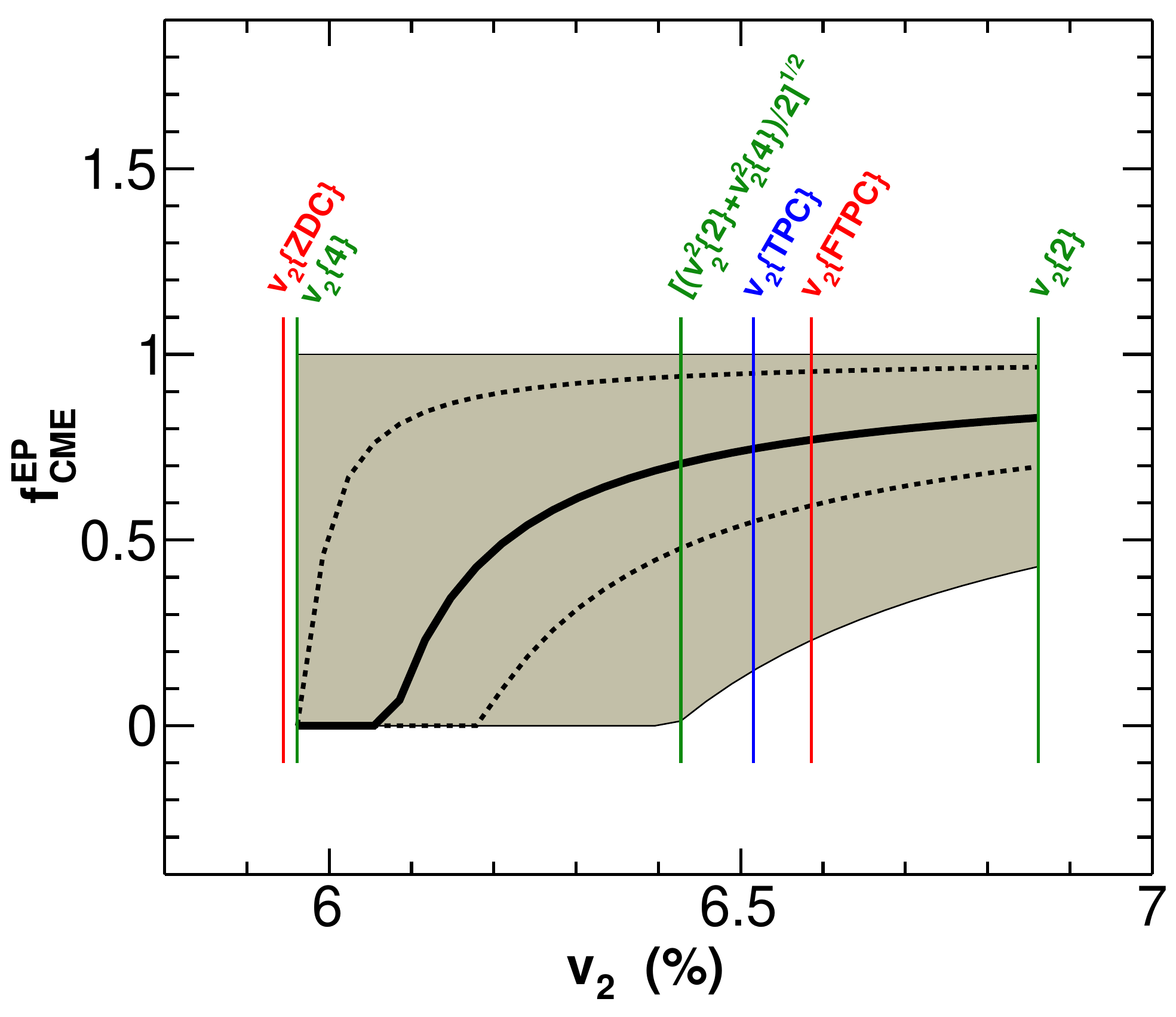}
  \caption{\label{fig:cme}
(Color online) Fraction of \cme\ contribution in the $\dgtwo$ measurement~\cite{Abelev:2009ac,Abelev:2009ad,Adamczyk:2013hsi} in the 20-60\% centrality range in 200A~GeV \AuAu\ collisions at \rhic\ versus ``true'' $\vv$. The gray area indicates the $\pm1\sigma$ statistical uncertainty, dominated by that in $\dgone$~\cite{Adamczyk:2013hsi}. The dashed curves would be the new $\pm1\sigma$ uncertainty with ten-fold increase in statistics.} 
    \end{center}
	\end{figure}

Table~\ref{tab:data} also lists the $\dg$ correlator measurements by STAR 
with respect to $\psitwo$~\cite{Abelev:2009ac,Abelev:2009ad,Adamczyk:2013hsi} 
and $\psione$~\cite{Adamczyk:2013hsi}. 
Although $\psione$ from \zdc\ may not strictly measure the \RP, our general formulism is still valid, and one can in principle extract the \cme\ signal from those $\dg$ measurements. 
Many of the experimental systematics related to event and track quality cuts cancel in their relative difference $\Rexp(\dg)\equiv2(\dgone-\dgtwo)/(\dgone-\dgtwo)$. The remaining major systematic uncertainty comes from those in the determinations of the \RP\ and \EP\ resolutions or the $\vv$~\cite{Abelev:2009ac,Abelev:2009ad}. 
In the STAR $\dgtwo$ measurement~\cite{Abelev:2009ac,Abelev:2009ad}, the $\vftpc$~\cite{Voloshin:2007af} was used and the systematic uncertainty was taken to be half the difference between $\vTT$ and $\vTF$. 
In the later STAR measurement~\cite{Adamczyk:2013hsi}, the $\dgtwo$ uncertainty is taken to be the difference between $\dgtwo$ and $\dgone$, perceived to be physically equal, but shown not to be the case by the present work. Below we use the later, higher statistics data~\cite{Adamczyk:2013hsi} but the earlier systematic uncertainty estimation~\cite{Abelev:2009ac,Abelev:2009ad}.
The systematic uncertainty was not estimated on $\dgone$~\cite{Adamczyk:2013hsi}, 
though statistical uncertainties are large and likely dominate. 
We average the $\vv$ and $\dg$ measurements over the centrality range 20-60\%, weighted by $\Npart^2$ (because the $\dg$ is a pair-wise average quantity).
We extract the \cme\ to \bkg\ ratio by Eq.~(\ref{eq:r}), replacing $\REP$ with $\Rexp$ and assuming $\aBEP=\av$, so $\Rexp(\Bsq)=-\Rexp(\vv)$. We vary the ``true'' $\vv$ over the wide range between $\vTT$ and $\vTF$, and at each $\vv$ the $\dgtwo$ is replaced by $\dgtwo\vftpc/\vv$ (i.e.~the three-particle correlator measurement divided by $\vv$). The fraction $\fcmeEP$ is obtained and shown in Fig~\ref{fig:cme} by the thick curve as a function of the ``true'' $\vv$. 
The gray area is the uncertainty, $\in[0,1]$, determined by the $\pm1\sigma$ statistical uncertainty in the $\dg$ measurements. The vertical lines indicate the various measured $\vv$ values. At present the data precision does not allow a meaningful constraint on $\fcmeEP$; the limitation comes from the $\dgone$ measurement which has an order of magnitude larger statistical uncertainty than that of $\dgtwo$. 
With ten-fold increase in statistics, the constraint would be the dashed curves. 
This is clearly where the future experimental emphasis should be placed: larger \AuAu\ data samples are being analyzed and more \AuAu\ statistics are to be accumulated; \zdc\ upgrade is ongoing in the CMS experiment at the \lhc; fixed target experiments at the SPS may be another viable venue where all spectator nucleons are measured in the \zdc\ allowing possibly a better determination of $\psione$.

\section{Summary}
In summary, elliptic flow ($\vv$) develops in relativistic heavy ion collisions from the anisotropic overlap geometry of the participant nucleons. The participant plane azimuthal angle ($\psiPP$), due to fluctuations, does not necessarily coincide with the reaction plane's ($\psiRP$). With respect to $\psiPP$, $\vv$ is stronger than that with respect to $\psiRP$. This has been known for over a decade.
The magnetic field (\Bbf) is, on the other hand, produced mainly by spectator protons and its direction fluctuates nominally about $\psiRP$, not $\psiPP$. Therefore, \Bbf\ with respect to $\psiPP$ is weaker than \Bbf\ with respect to $\psiRP$. This has so far not been well appreciated.
We have verified these with MC Glauber (\mcg) calculations and A Multi-Phase Transport (\ampt) model simulations of \AuAu, \CuCu, \RuRu, \ZrZr, and \PbPb\ collisions. 
One can effectively ``change'' \Bbf\ in a single nucleus-nucleus collision and, at the same time, ``change'' $\vv$ in the opposite direction; the change is significant, as large as 20\% in each direction in \AuAu\ collisions.
We demonstrate that this opposite behavior in a single collision system, thus with small systematic uncertainties, can be exploited to effectively disentangle the possible chiral magnetic effect (\cme) from the $\vv$-induced background in three-point correlator ($\dg$) measurements.
We argue that the comparative measurements of $\dg$ with respect to $\psiRP$ and $\psiPP$ in the same collision system is superior to isobaric collisions where large systematics persist.

We have applied this novel idea to experimental data; however, due to poor statistical precision of the data, no conclusion can presently be drawn regarding the possible \cme\ magnitude. 
This calls for future efforts to accumulate data statistics and to improve capabilities of Zero-Degree Calorimeters.
With improved statistics, the novel method we report here should be able to decisively answer the question of the CME in quantum chromodynamics.

\section*{Acknowledgments}
This work was supported in part by the National Natural Science Foundation of China under Grants No.~11647306, 11747312, U1732138, 11505056, 11605054, and 11628508, and US~Department of Energy under Grant No.~DE-SC0012910.

\bibliographystyle{unsrt}
\bibliography{../../ref}
\end{document}